\documentclass[preprint,aps,floats,floatsfix,showpacs]{revtex4-1}

\usepackage{dcolumn} 
\usepackage{bm} 
\usepackage{amsthm}
\usepackage{latexsym}
\usepackage{float}
\usepackage{amsfonts}
\usepackage{amsmath}
\usepackage{amssymb}
\usepackage{color,graphicx}

\begin{document}

\title{Electronic Coherence Control in a Charged Quantum Dot}

\author{G.~Moody}
\affiliation{National Institute of Standards and Technology, Boulder CO 80305}
\author{C.~McDonald}
\affiliation{National Institute of Standards and Technology, Boulder CO 80305}
\author{A.~Feldman}
\affiliation{National Institute of Standards and Technology, Boulder CO 80305}
\author{T.~Harvey}
\affiliation{National Institute of Standards and Technology, Boulder CO 80305}
\author{R.~P.~Mirin}
\affiliation{National Institute of Standards and Technology, Boulder CO 80305}
\author{K.~L.~Silverman}
\email{kevin.silverman@nist.gov}
\affiliation{National Institute of Standards and Technology, Boulder CO 80305}

\begin{abstract}
 Minimizing decoherence due to coupling of a quantum system to its fluctuating environment is at the forefront of quantum information science and photonics research. Nature sets the ultimate limit, however, given by the strength of the system's coupling to the electromagnetic field. Here, we establish the ability to electronically control this coupling and \textit{enhance} the coherence time of a quantum dot excitonic state. Coherence control is demonstrated on the positively charged exciton transition (an electron Coulomb-bound with two holes) in quantum dots embedded in a photonic waveguide by manipulating the electron and hole wavefunctions through an applied lateral electric field. With increasing field up to 15 kV cm$^{-1}$, the coherence time increases by a factor of two from $\sim1.4$ ns to $\sim2.7$ ns. Numerical calculations reveal that longer coherence arises from the separation of charge carriers by up to $\sim6$ nm, which leads to a $30\%$ weaker transition dipole moment. The ability to electrostatically control the coherence time and transition dipole moment opens new avenues for quantum communication and novel coupling schemes between distant qubits.
\end{abstract}

\date{\today}
\maketitle

\begin{figure}[h]
\centering
\includegraphics[width=0.6\columnwidth]{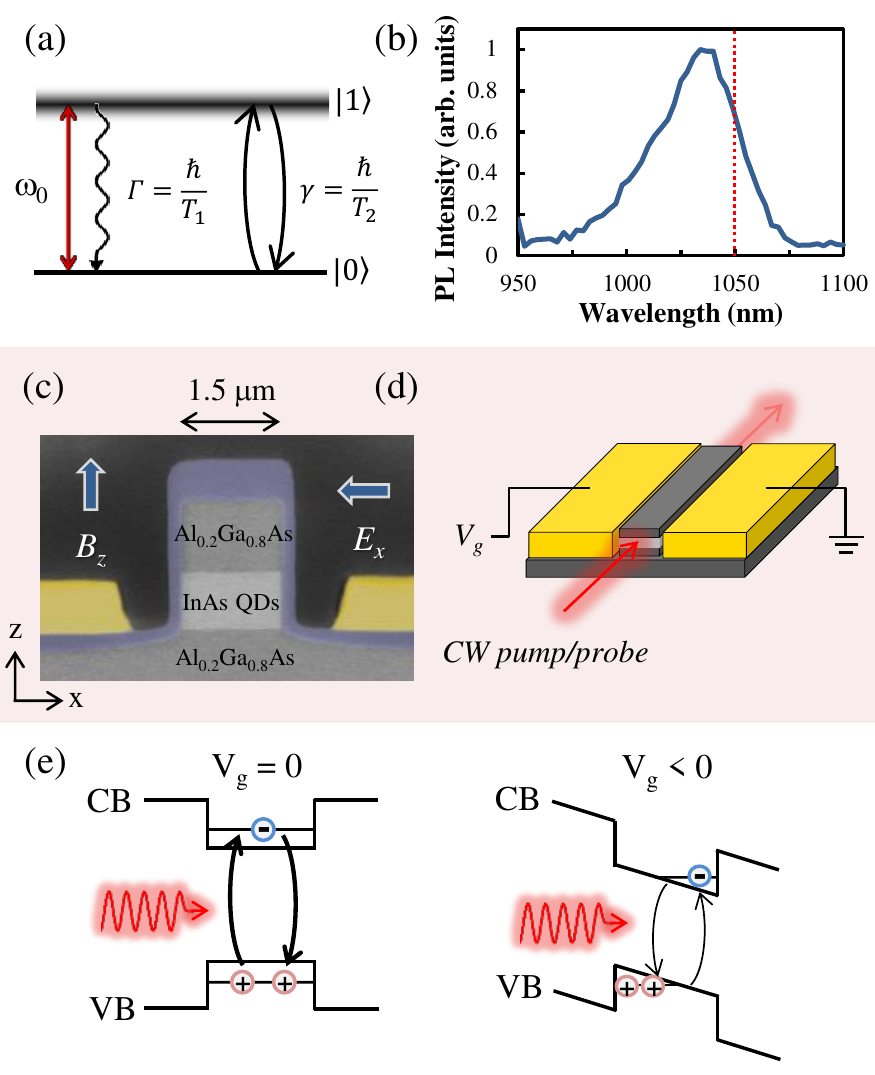}
\caption{(a) The quantum dynamics of a simple two-level system with resonance frequency $\omega_0$ are governed by radiative recombination with rate $\Gamma$ (radiative lifetime $T_1$) and decoherence with rate $\gamma$ (coherence time $T_2$), which defines the homogeneous linewidth. (b) Photoluminescence spectrum taken at 4.2 K showing the inhomogeneously broadened ground state trion transition. The laser wavelength at 1050 nm is indicated by the vertical red dashed line. (c) False color scanning electron microscope image of the QD waveguide device and direction of the applied electric (\textbf{$E_X$}) and magnetic (\textbf{$B_Z$}) fields. (d) Continuous wave (CW) pump and probe lasers are collinearly coupled into the waveguide for the spectral hole burning measurements. Gold electrodes are patterned on both sides of the waveguide for the application of a lateral electric field. (e) Overlap of the electron and hole wavefunctions results in strong optical absorption and emission in the absence of an external electric field. Application of a lateral ($x$-direction) electric field separates the electron and hole wavefunctions in the QD, reducing their overlap integral. A smaller transition dipole moment results in weak radiative coupling, increasing the radiative lifetime.}
\label{fig1}
\end{figure}

In the solid state, three-dimensional quantum confinement of charge carriers in a semiconductor quantum dot (QD) decouples them from their surroundings, resulting in robust optical coherence of the exciton (Coulomb-bound electron-hole pairs) and trion (excitons bound to an additional electron or hole) states.\cite{Borri2001b} Excitonic coherence in QDs is a fundamental property of light-matter interaction and plays an important role in opto-electronics. From a quantum information perspective, the coherence time ($T_2$) is a key parameter for quantum phenomena including the duration of Rabi oscillations,\cite{Press2008} fidelity of spin-photon entanglement,\cite{Gao2012} cavity-emitter coupling,\cite{Rakher2009} and photon indistinguishability.\cite{Kiraz2004,Flagg2011} The leading source of exciton and trion decoherence at elevated temperatures is coupling of charge carriers to phonons, which destroys coherence on a picosecond timescale. At cryogenic temperatures where electron-phonon scattering is absent, a one-nanosecond $T_2$ time has been measured corresponding to a sub-$\mu$eV homogeneous linewidth $\gamma$ (inversely proportional to $T_2$, see Fig. \ref{fig1}(a)), which is limited primarily by the excited-state recombination lifetime.\cite{Berry2006b,Cesari2010b,Moody2014}

Fast and deterministic control of $T_2$ would be an enabling technology for future semiconductor QD photonic devices. Simply increasing $T_2$ would extend the time available for coherent rotations of electronic states about the Bloch sphere. With careful control one can optimize the trade off between fast rotation times to suit the properties of the source and longevity of coherence for robust operations. The coherence time also sets the extent of the single-photon wavepacket, which can be leveraged to improve the purity and indistinguishability of single photons generated from unique QDs for linear optical quantum computing applications.\cite{Santori2004} With control of the coherence time faster than $1/T_2$, more advanced operations such as dynamic tunability of the Rabi frequency and coherent storage of QD qubits can be envisioned.

Despite the above-mentioned utility, control of excitonic coherence in QDs has not been intensely explored. This is especially true when it comes to increasing the coherence time. A large body of work is devoted to changing the spontaneous emission lifetime ($T_1$) of QDs by embedding them in nanostructures with a modified vacuum density of optical states. Large enhancements of the emission rate can be achieved via the Purcell effect,\cite{Englund2005,Reithmaier2004,Hennessy2007} and if $T_1$ becomes smaller than $T_2/2$, the coherence time will \textit{decrease} by further reducing $T_1$. The radiative lifetime can also be increased in these structures, but this has no effect on the coherence time as it is already limited by pure dephasing in these cases.

Alternatively, one can control the spontaneous emission rate by manipulating the electron and hole wavefunctions within the QD using an external electric field.\cite{Robinson2005,Gerardot2007,Vogel2007,Alen2007,Jarjour2007} While in many of these works the motivation was to tune the exciton fine-structure splitting, modifying the electron-hole overlap also affected the homogeneous linewidth and emission lifetime. In the majority of previous studies, carrier tunneling out of the QDs tended to decrease the lifetime and broaden the linewidth; however the introduction of a high-energy barrier to suppress tunneling led to a factor of two increase in the lifetime for vertically \cite{Bennett2010} and laterally \cite{Reuter2006b} applied electric fields aligned along the growth and in-plane directions, respectively. Despite these advancements in spontaneous emission control, lifetime measurements do not provide any details of the coherent properties of the system. In many experiments it was extremely likely $T_2$ was limited by non-radiative recombination and pure dephasing due to background charges and non-resonant excitation of carriers.

In this work we demonstrate a new approach for electronically controlling and \textit{enhancing} excitonic coherence in QDs. We examine a single layer of charged InAs/GaAs QDs emitting near 1045 nm with a $\sim50$ nm inhomogeneous linewidth due to QD size and composition fluctuations (see Fig. \ref{fig1}(b) and Supplementary Information). A photonic ridge waveguide structure is lithographically defined to confine optical modes to the QD region, illustrated by the false color scanning electron microscope image in Fig. \ref{fig1}(c). Gold electrodes are patterned on both sides of the waveguide. Applying a quasi-static voltage bias across the electrodes generates an approximate in-plane electric field up to 15 kV cm$^{-1}$ at the QD region. The applied field depicted in Fig. \ref{fig1}(e) tilts the conduction (CB) and valence (VB) bands in the plane of the device along the $x$-direction. Upon optical excitation of the trion transition, lateral separation of the electron and holes decreases the spatial overlap of their wavefunctions, reducing the radiative efficiency and increasing the radiative lifetime. Using high-resolution optical spectroscopy, we demonstrate that coherence is maintained during this process and the homogeneous linewidth (coherence time) decreases (increases) by a factor of two from $\gamma = 0.45$ $\mu$eV to $\gamma = 0.2$ $\mu$eV. Since our QDs exhibit a Fourier-limited emission profile under these experimental conditions, we have full control of the coherence time through its fixed relationship to the radiative lifetime.

This device design offers several advantages for electronic coherence control compared to previous studies. First, the photonic ridge waveguide structure confines the optical modes to the QD region, enhancing the light-matter interaction. Second, a lateral-field geometry is chosen for manipulating the electron and hole wavefunctions because the base diameter of our QDs is much larger than the height. Compared to diode devices generating longitudinal fields along the sample growth direction, a lateral field can generate much larger separation between charges before becoming strong enough to induce significant tunneling out of the QD and thus destruction of coherence. The structure also has extremely low capacitance ($\sim100$ fF), offering the possiblity of gigahertz frequency control of the coherence time.

The trion coherence time is obtained by measuring the homogeneous linewidth using spectral hole-burning spectroscopy (see Fig. \ref{fig1}(d) and Supplementary Information). Briefly, continuous-wave pump and probe lasers are collinearly coupled into the QD waveguide device, held at 4.2 K in a transmission confocal microscopy setup. The pump laser wavelength is fixed at 1050 nm and partially saturates the trion absorption, burning a spectral ``hole'' in the inhomogeneous distribution. A probe laser is spectrally tuned through the pump to map out the lineshape of the spectral ``hole,'' providing a measure of the homogeneous linewidth of the trion transition. This technique allows for shot-noise limited detection of pump-induced modulation of the probe absorption and is insensitive to low frequency spectral diffusion. The sample preparation and geometry enable our electronic coherence control.

\begin{figure}[h]
\centering
\includegraphics[width=0.8\columnwidth]{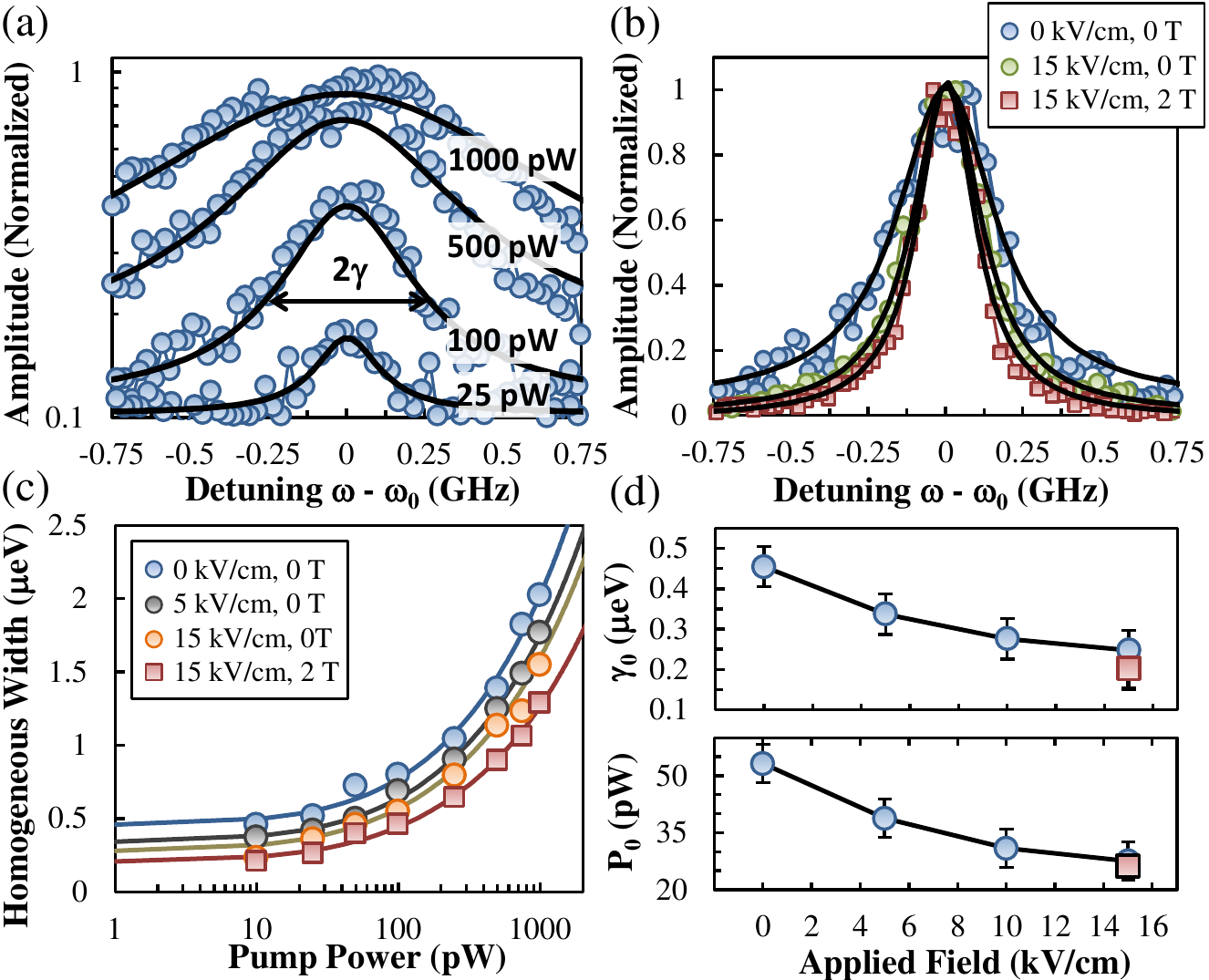}
\caption{(a) Homogeneous lineshape as a function of pump laser power from 25 pW to 1000 pW (points) and Lorentzian fit functions (solid lines). The half-width at half-maximum provides a measure of the homogeneous linewidth, $\gamma$. (b) Homogeneous lineshape for 100 pW pump power and an applied lateral field of 0 kV cm$^{-1}$ (blue circles), 15 kV cm$^{-1}$ (green circles), and 15 kV cm$^{-1}$ in addition to a 2 T magnetic field in Faraday geometry (red squares). The solid lines are Lorentzian fit functions. (c) Power broadening of the homogeneous linewidth (symbols) for applied field up to 15 kV cm$^{-1}$. A fit of Eqn. \ref{eqn1} (solid lines) reveals that both the zero-power linewidth, $\gamma_0$, and saturation power, $P_0$, decrease with increasing field, shown in top and bottom panels in (d), respectively. The square data points are in the presence of a 2 T magnetic field in Faraday configuration.}
\label{fig2}
\end{figure}

We first present measurements with zero applied field. Homogeneous lineshapes are shown in Fig. \ref{fig2}(a) (solid points) for increasing pump power from 25 pW to 1 nW. The half-width at half-maximum (HWHM) of the Lorentzian fit function (solid lines) provides a measure of the homogeneous linewidth at each power and is inversely proportional to the coherence time.\cite{EndNote1,Sargent1978} The linewidth increases with pump power due to saturation broadening of the transition (blue symbols in Fig. \ref{fig2}(c)), which is consistent with the expected behavior for a resonantly driven, isolated two-level system. Saturation broadening of the linewidth can be expressed as

\begin{equation}
\gamma = \frac{\gamma_0}{2}\sqrt{1+P/P_0},
\label{eqn1}
\end{equation}

where $P$ is the pump power, $P_0$  is the pump power required to saturate the absorption, and $\gamma_0$ is the transition linewidth in the absence of saturation effects.\cite{Allen1987} The zero-power limit of the fit (solid blue line in Fig. \ref{fig2}(c)) is used to determine a linewidth of $\gamma_0 = 0.45\pm0.05$ $\mu$eV ($T_2 = 1.46\pm0.07$ ns). This value is consistent with population-lifetime limited results previously measured for QD trions, indicating that additional pure dephasing mechanisms such as electron-phonon coupling and spectral diffusion effects are absent on the measurement timescale.\cite{Moody2014,Cesari2010b} From the same fit we also extract the absorption saturation power $P_0 = 53\pm5$ pW. Since a precise determination of the pump power coupled to the QDs in the waveguide is difficult, this value only provides an estimate of this fundamental property of light-matter interaction. Nonetheless, relative changes of $P_0$ contain useful information on how easily the QD absorption saturates under different experimental conditions.

Our major results are illustrated by the homogeneous lineshapes presented in Fig. \ref{fig2}(b) for 100 pW pump power. Upon applying a bias to the electrodes, the homogeneous lineshape exhibits an electric-field-induced narrowing. This behavior is present for all incident pump powers, shown in Fig. \ref{fig2}(c) for fields up to 15 kV cm$^{-1}$. The linewidth exhibits saturation broadening behavior at all applied field strengths, which is fit using Eqn. \ref{eqn1} to determine how the lateral field influences $\gamma_0$ and $P_0$. Both the homogeneous linewidth and saturation power decrease monotonically to $\gamma_0 = 0.25\pm0.05$ $\mu$eV and $P_0 = 28\pm5$ pW at 15 kV cm$^{-1}$ (Fig. \ref{fig2}(d)). At larger field strengths, the linewidth increases suggesting the onset of charge carrier tunneling out of the QDs. Although the lineshape is predominantly governed by radiative recombination, previous work has shown that precession of the trion-bound electron spin about the Overhauser magnetic field created by the $\sim10^5$ nuclear spins in the QD results in additional slow decoherence.\cite{Moody2014} This electron hyperfine-mediated broadening can be screened by a moderate magnetic field in the Faraday configuration, which is confirmed here by additional linewidth narrowing to $\gamma_0 = 0.20\pm0.05$ $\mu$eV ($T_2 = 3.29\pm0.07$ ns) for a magnetic field strength of 2 T (red squares in Fig. \ref{fig2}(d)). Along with the ability to electrically control the coherence time, the linewidth of 0.20 $\mu$eV is the narrowest reported for group III-V QDs to date. We focus on the changes in the lineshape here because it is the least sensitive to systematic errors, but fits to the absolute change in absorption yield consistent results and are shown in the Supplementary Information.

\begin{figure}[t]
\centering
\includegraphics[width=0.6\columnwidth]{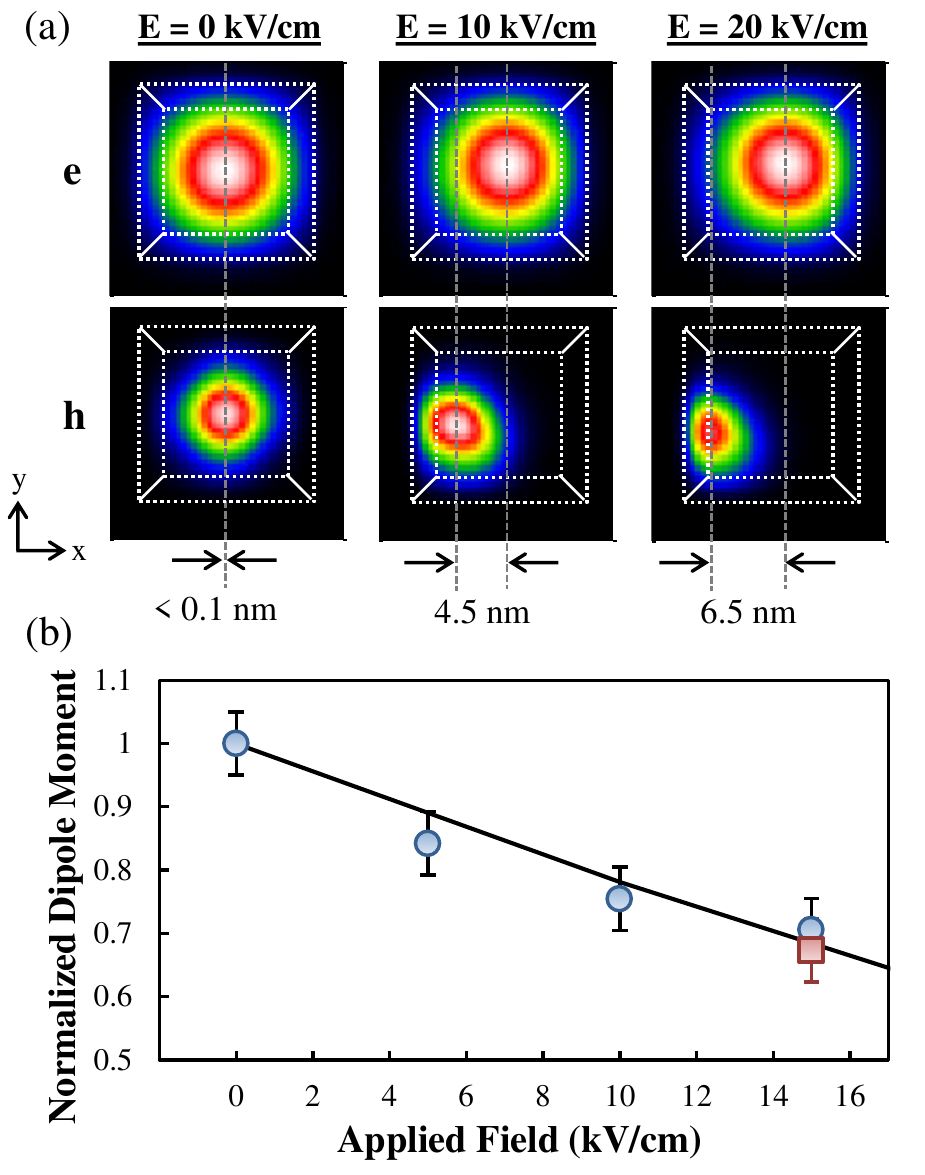}
\caption{(a) Single-band effective-mass calculations of the single-electron (top row) and two-hole (bottom row) wavefunctions as a function of applied lateral electric field. The lateral field pulls the electron and holes in opposite directions by up to $\sim 6$ nm before tunneling destroys the coherence. (b) Calculated (solid line) and measured (symbols) transition dipole moment as a function of applied electric field. The square data point is in the presence of a 2 T magnetic field in Faraday configuration.}
\label{fig3}
\end{figure}

These results clearly demonstrate that the applied field decreases the trion transition dipole moment, consequently increasing the coherence time. To illustrate this, we can express the homogeneous linewidth and saturation power in terms of the radiative lifetime ($T_1$), coherence time ($T_2$), and transition dipole moment ($\mu$) through the relations $\gamma_0=1/2\left(\Gamma+\Gamma'\right)$ and $P_0 \propto \left(T_1T_2\mu^2\right)^{-1}$, where $\Gamma' = 0.05$ $\mu$eV is equal to the measured hyperfine interaction broadening in this sample.\cite{Moody2014,Allen1987} Additionally, in semiconductor QDs the radiative lifetime can be related to the transition dipole moment through $T_1 \propto \mu^{-2}$. After inserting these expressions into Eqn. \ref{eqn1} we obtain

\begin{equation}
\gamma = \frac{1}{4}\left(\mu^2 + \Gamma'\right)\sqrt{1+\frac{P}{2\left(\mu^2+\Gamma'\right)}},
\label{eqn2}
\end{equation}
which is a simple expression relating the linewidth power broadening to relative changes in the transition dipole moment. We show in Fig. \ref{fig3}(b) (points) the normalized transition dipole moment extracted from the linewidth fits as a function of applied electric field, normalized to the zero-field value. With increasing field up to 15 kV cm$^{-1}$, the dipole moment decreases by 30 $\%$, which is reflected in the field-induced reduction in the linewidth and saturation power in Fig. \ref{fig2}(d). Interestingly, screening of the hyperfine interaction by an external magnetic field results in a narrower linewidth without affecting the transition dipole moment (square symbol in Fig. \ref{fig3}(b)), $\it{i.e.}$ the hyperfine-mediated broadening introduces additional decoherence to the system without reducing the radiative efficiency. This result is consistent with our measurements of similar saturation powers with and without the Faraday magnetic field shown in Fig. \ref{fig2}(d).

We rule out a quantum-confined Stark shift\cite{Gotoh1997,Gotoh2000,Seufert2001} as a source of linewidth narrowing, which could shift QDs with a smaller linewidth into resonance with the pump, by measuring homogeneous lineshapes for QD emission at 1045 nm (see Supplementary Information). We obtain $\gamma = 0.50\pm0.05$ $\mu$eV at 0 kV cm$^{-1}$ and 25 pW pump power, equal to the value obtained when tuned to 1050 nm within the estimated uncertainty ($0.52\pm0.05$ $\mu$eV). Equivalent linewidth values are also measured for a 1 nW pump power ($2.03\pm0.05$ $\mu$eV and $2.11\pm0.05$ $\mu$eV at 1050 nm and 1045 nm, respectively). We therefore attribute the decrease in the transition dipole moment, leading to a longer coherence time, to a large in-plane dipole moment created by the applied electric field.

Quantitative insight into field effects on the electron and hole wavefunctions is obtained through numerical calculations using a single-band effective-mass Hamiltonian including a full treatment of piezoelectric effects and strain.\cite{nextnano,Pryor1998} We consider QDs with uniform composition profiles that are square-based truncated pyramids with a 20 nm base and 5 nm height, similar to the average QD size measured with atomic force microscopy (see Supplementary Information). In the top and bottom panels of Fig. \ref{fig3}(a) we show normalized in-plane single electron and two-hole wavefunction probabilities. For increasing field strength along the $x$-direction up to 20 kV cm$^{-1}$, the electron and two-hole wavefunctions are shifted in opposite directions, with the holes shifted to lower values along the $y$-direction due to the built-in piezoelectric charges. An applied field up to 15 kV cm$^{-1}$ leads to an in-plane dipole moment per unit charge of $\sim6$ nm. The electron-hole separation decreases the overlap integral of the wavefunctions resulting in a weaker transition dipole moment with increasing applied field (solid line, Fig. \ref{fig3}(b)), which nicely reproduces the measurements (symbols). The calculated lateral dipole moment is nearly an order of magnitude larger than the permanent dipole moment measured in vertical field-effect devices,\cite{Finley2004} demonstrating the advantage of the lateral field geometry for wavefunction manipulation in QDs. Wavefunction calculations were also performed for a single electron and single hole to model electric field effects on the exciton dipole moment in neutral QDs (Supplementary Information). Qualitatively similar trends for the electron-hole separation and overlap integral are obtained.

Electronic control of the excitonic dipole moment and coherence time provides new opportunities for integrated QD photonics and quantum information processing. For example, quantum operations of electronic spin qubits can be performed with the trion transition acting as an intermediary state. The ability to tune the trion transition dipole moment provides an additional control knob for rotating the electronic spin about the Bloch sphere. The large transition dipole moment also makes QDs particularly attractive as efficient non-classical light sources. Scaling beyond a single QD for quantum computing and communication requires that different QDs generate indistinguishable photons,\cite{Bennett2009} which may benefit from the Fourier-limited coherence control demonstrated here. Moreover, the large field-induced, in-plane static dipole moment might facilitate scalable two-qubit quantum logic gates through the dipole-dipole interaction. Previous work has demonstrated weak dipolar coupling ($\sim30$ $\mu$eV) between neighboring QDs, the strength of which depends on the dipole moment squared.\cite{Unold2005b} The calculated increase in the lateral dipole moment by over an order of magnitude with applied field implies a more than one-hundred-fold increase in the dipolar coupling strength--sufficient to implement nonlocal conditional quantum logic.\cite{Biolatti2000} These effects could be enhanced by tailoring the QD size, shape, and composition to optimize the confinement potential for maximal separation of electron and hole wavefunctions. We anticipate the results presented here will motivate additional studies investigating novel solid-state coherent control and coupling schemes using semiconductor QDs.

\bibliography{RefList1}{}

\begin{thebibliography}{32}
\bibitem [{End()}]{EndNote1}%
  \BibitemOpen
  \href@noop {} {\bibinfo  {journal} {The single Lorentzian lineshape indicates
  that pure dephasing processes are absent from the spectra and the homogeneous
  linewidth is limited by the population lifetime. Contributions to the
  nonlinear signal from spectral-hole burning and coherent population
  oscillations spectrally overlap with similar linewidths.}\ }\BibitemShut
  {NoStop}%
\end{thebibliography}

\end{document}